\documentclass[english]{aa}
\usepackage{txfonts}
\usepackage{float}
\usepackage{graphicx}

\providecommand{\tabularnewline}{\\}

\usepackage{natbib}
\usepackage{graphicx}
\usepackage{babel}
\begin{document}

\title{The Distance to NGC 5904 (M 5) via the Subdwarfs Main Sequence Fitting
Method \thanks{Based on data collected at ESO-La Silla, Chile, (GTO 63.L-0717)
  and from HST archival data (GO 8310)} }

\author{V. Testa\inst{1},A. Chieffi\inst{2}, M.Limongi\inst{1},
  G. Andreuzzi\inst{1,3}, and G. Marconi\inst{1,4}}

\institute{INAF-OAR, Via Frascati 33, 00040 Monteporzio Catone ITALY \and
INAF-IASF, Via del Fosso del Cavaliere, 00100, Roma ITALY \and
INAF-TNG, Santa Cruz de La Palma, Canary Islands, Spain \and
ESO-Chile, Alonso de Cordova 3107, Vitacura, Casilla 19001, Santiago 19, CHILE}

\offprints{Vincenzo Testa\email{testa@mporzio.astro.it}}

\date{received 13 Oct 2003/accepted 04 Mar 2004}

\abstract{We present a determination of the distance modulus of the globular
cluster NGC 5904 (M 5), obtained by means of the subdwarf
main--sequence fitting on the (V,V$-$I) color--magnitude
diagram. The subdwarf sample has been selected from the HIPPARCOS
catalog in a metallicity range homogeneous with the cluster ([Fe/H]
$\simeq$ $-$1.1). 
Both the cluster and the subdwarfs have been observed with the
same telescope+instrument+filters setup (namely, ESO--NTT equipped
with the SUSI2 camera), in order to preserve
homogeneity and reduce systematic uncertainties. A set of archival HST
data has then been used to obtain a deep and precise ridge
line. These have been accurately calibrated in the ground photometric
system by using the NTT data and used to fit the cluster distance
modulus. By adopting the most commonly accepted values for the
reddening, E(B$-$V) = 0.035 and 0.03, we obtain respectively $\mu_0
= 14.44 \pm 0.09 \pm 0.07$ and $\mu_0 = 14.41 \pm 0.09 \pm 0.07$, 
in agreement with recent determinations.

}

\authorrunning{Testa et al.}
\titlerunning{Distance to M 5}

\maketitle

\section{Introduction}

The ages of the Galactic Globular Clusters (hereafter GGCs) represent
one of the milestones on the road of our comprehension of the Universe.
In order to derive reliable absolute ages estimates, the most important
parameter to determine is the distance \citep[see][]{renz91}. Errors
on the distance propagates through the age determination procedure
in such a way that a 0.08 error in the distance modulus (DM) 
propagates as a 12\% error
in the age. Different ages (younger or older) mean different
epoch of formation of the Galaxy. Therefore, it can be easily seen
how the uncertainty in GGCs ages estimates significantly affects models
of galaxy formation and cosmological evolution. 

The recent availability of the \textit{HIPPARCOS} catalog of trigonometric
parallaxes \citep{perryman97} has greatly improved the knowledge of GGCs distances.
A large number of papers on the subject have been published since
then \citep[e.g.][]{reid97,gratton97,pont98,vdb02,gratton03}, most of which present
larger distances, and hence younger ages, with respect to previous
works \citep[with the exception, among the cited works, of][who recover the
pre-Hipparcos results of 1992]{pont98}.

The method used consists of a fit of the cluster main sequence to
the subdwarfs main-sequence of corresponding metallicity, once the
reddening has been subtracted. This procedure is, of course, prone to uncertainties,
both random and systematic: on the subdwarf side errors on parallax
measurements, Lutz-Kelker correction, metallicity; on the cluster
side reddening, metallicity, narrowness and accuracy of the mean ridge
line. A further effect is represented by the presence of binary stars in the
cluster CMD and in the subdwarfs sample. In the first case the determination
of the mean ridge line is altered, if a proper fitting procedure is not taken into
account. In the second, the magnitude of the field star is different from 
stars of analogous metallicity. This
issue has been widely discussed in, e.g., \cite{gratton97,pont98,carretta00}, 
and will be analyzed in detail later.
If \textit{HIPPARCOS} gave the chance to significantly reduce
the random errors on parallax measurements, nonetheless other sources
of errors still remain. Metallicity, for example, represents a common
problem that significantly affects the determination of the distance
modulus, because of either the spread in metallicity of the sequence
of local subdwarfs or the error on the absolute values. There are
two strategies to go around this problem: to obtain a sufficiently
large set of subdwarfs in the right metallicity range, or to apply
a relation that transforms the colors of the stars of different metallicity.
The second introduces a model-dependent relation and should be used
with caution. Of course, the first solution is better, especially
if the sample of subdwarfs draws a main sequence down to very low
masses. In fact, recent HST color-magnitude diagrams of GGCs 
reach very faint magnitudes, hence small masses, but a similar sequence
for subdwarfs is currently not available. \cite{reid98} fit a distance
modulus to NGC 6397 by using a sequence of lower MS subdwarfs, but
their work stands on the fact that the sequence of very metal-poor
subdwarfs define a reasonably narrow sequence. This is not true, instead,
for subdwarfs of higher metallicities (see their Fig. 2). On the other
hand, the most widely used set of subdwarfs cover magnitudes ranging
from $\mathrm{M}_{V}\sim7$ and brighter, leaving uncovered the range between
this value and the extreme subdwarfs. This is explained with the relative
lack of halo stars near the Sun \citep{reid98}. One of the possibilities
we discuss in order to reduce the uncertainties is to perform a uniform
study of a subdwarfs sample and a GGC by using the same telescope,
instrumentation and, possibly, observing nights, thus eliminating the systematics
due to the relative calibrations among different instrumental setups.
For this reason, we observed the target cluster and a sample of subdwarfs
in the proper metallicity range matching the cluster's with the same
telescope+instrument (NTT+SUSI2) and then used archival data from
HST to go faint and calibrated the HST magnitudes into the ground
system, obtaining a deep and fully homogenous data set. The selected
object is the well known cluster NGC 5904 (M 5), for which a wide
set of photometric and spectroscopic studies is available in the literature.
Moreover, \citet[S96]{sandquist96}, \cite{reid97}, \cite{gratton97},
\cite{chaboyer98} and, more recently, \cite{vdb02} determined the 
DM of the cluster by using the trigonometric parallaxes of
subdwarfs and various sets of models. These works will be often referred to in
the following. 

NGC 5904 is a typical old halo cluster of intermediate
metallicity -{[}Fe/H{]} $\sim -1.1$ in the scale of \cite{cg97} (CG),
$-$1.4 in the \cite{zw84} scale (ZW). A recent study of \cite{ramirez03}
gives {[}Fe/H{]} $\simeq -1.3$. The morphology of the horizontal
branch (HB) does not show signs of the second parameter effect and
several age determinations are available. The structure of the
paper is outlined as follows: Sect. 2 describes the observations and
the archival material, Sect. 3 presents the reduction and calibration
of the data, in Sect. 4 we present and analyze the subdwarfs sample. In
Sect. 5 the cluster color-magnitude diagram (CMD) is presented and
discussed, and in Sect. 6 the DM-fit is analyzed. In Sect. 7 a final
discussion is summarized.

\section{Data Set}

\subsection{Observations}

Observations have been performed with the ESO-NTT telescope equipped
with SUSI2 as a backup program during three different campaigns, 
in december 1999 (63.L-0717, subdwarfs + cluster), 
december 2000 (66.D-0242, subdwarfs), 
february 2002 (68.B-0061, cluster). 
The adopted strategy was to observe both the subdwarfs 
sample and the target cluster with the same setup
telescope+instrument+filters+detector 
and, possibly, within the same campaign, 
in order to remove any systematics due to the
use of different setups. The december 1999 and february 2002 campaigns
produced very good quality data, while the december 2000 set (two
subdwarfs only) presented moderately poor photometric conditions.
Seeing was not a major problem for observing the subdwarfs,because
they had to be defocussed in order not to saturate the chip. This, coupled to
the fact that the objects were observed in the central region of the frames, 
allowed to use exposure times long enough to have a negligible shutter effect. 
The typical observing time, per frame, was 0.5 s. 

The data
set for the target cluster (NGC~5904 = M~5) was secured in december
1999, when the seeing was poor, together with the subdwarfs, and in
february 2002 when seeing was considerably better, in order to have
a comparison data-set. In both cases, the run was found to be photometric
and, in fact, the run-to-run scatter in the magnitudes is almost absent,
without any trend in the magnitudes and colors (see Fig \ref{figruntorun}).
The field was selected 2 arcmin North of the cluster center, to ensure
an almost complete overlapping with the field observed with HST that
have been retrieved from the archive (see below). Only images in V
and I bands for the cluster were taken, because the space-based data-set
consists of images in these two bands only. For the subdwarfs, also
B frames were secured. 

\subsection{Archival data}

A set of deep exposures was available from the STScI HST WFPC2 archive
(GO 8310), in the two wide bands {\small F555W} (V) and {\small F814W}
(I). A total of 6 {\small F555W} and 9 {\small F814W} frames were
retrieved from the archive, and pre-reduced by using the HST pipeline,
available in STSDAS running under IRAF \citep{tody86,tody93}\footnote{IRAF is distributed by 
the National Optical Astronomy Observatories, which are operated by the 
Association of Universities for Research in Astronomy, Inc., under 
cooperative agreement with the National Science Foundation.},
and using the most recent calibration frames and tables (see the
WFPC2 web page for details).

\section{Reductions and Calibrations}

\subsection{Reductions}

Reductions have been done for all the data-sets by using IRAF and
the \textit{digiphot} packages \textit{daophot} and \textit{photcal}
for building up the catalog and apply photometric calibrations respectively.
In particular, since the telescope had been defocused when observing
the subdwarfs sample, for these stars aperture photometry has been
used. For cluster images, both NTT and HST, we applied the standard
daophot PSF fitting routines. In particular, the NTT data-set is characterized
by a relatively high oversampling (FWHM $\sim$ 5 pix), and in this
case a \textit{gaussian} model for the PSF plus look-up table of correction
varying quadratically along the frames has been used. HST-WFPC2 frames,
on the other hand, are undersampled (estimated FWHM $\sim$ 1.6 pix),
and in this case a \textit{moffat15} function, with quadratic look-up
table, gave better results. As a general strategy, object searching
has been done independently on each frame, after registering all of
them to a common reference frame (the best seeing one). In the case
of HST, all the images were almost perfectly registered onto each
other, and no shift or rotation was needed. We made use of the world
coordinate system available in the FITS header in order to obtain
later an astrometric calibration of the catalogs.

\subsection{Photometric calibrations}

For both data-set, aperture corrections were applied in order to report
the PSF magnitudes to aperture magnitudes before calibrating. HST-WFPC2
data were reported to the standard aperture of 0.1$^{\arcsec}$, following
the calibration procedure of \citep{holtzmann95}. For ESO-NTT data,
growth-curves (from DAOGROW) have been used and aperture corrections
at infinity have been computed. For ground-based data, a set of Landolt
standards has been secured for each run and calibration equations
derived, in the form \ref{eq1}. Table \ref{calcoeff} reports the
calibration coefficients of all the runs. The equation for the V filter
has been derived against $\mathrm{B-V}$ in the two first runs, as B,V,I data
were available both for subdwarfs and for NGC 5904. In the last run,
only NGC 5904 has been observed and only in V and I filters, hence
we calibrated the V magnitude against (V-I).

\begin{eqnarray}
\label{eq1}
v & = & V+a_{0}+a_{1}\times(b-v)+a_{2}\times X_{V}\nonumber \\
v & = & V+a_{0}^{\prime}+a_{1}^{\prime}\times(v-i)+a_{2}^{\prime}\times X_{V}\\
b & = & B+b_{0}+b_{1}\times(b-v)+b_{2}\times X_{B}\nonumber \\
i & = & I+c_{0}+c_{1}\times(v-i)+c_{2}\times X_{I}\nonumber 
\end{eqnarray}

\begin{table}[htbp]

\caption{Calibration coefficients for the three runs.}

\label{calcoeff}

\begin{tabular}{cccc}
\hline 
Filter&
 Zero Point&
 Color Term&
 Extinction\tabularnewline
\hline
\multicolumn{3}{c}{ \textbf{Dec. 1999}}&
\tabularnewline
 B &
 25.42(0.01) &
 -0.05(0.01) &
 0.23(0.01) \tabularnewline
 V &
 25.82(0.01) &
 0.02(0.01) &
 0.11(0.01) \tabularnewline
 I &
 24.74(0.01) &
 -0.04(0.01) &
 -0.04(0.01) \tabularnewline
\multicolumn{3}{c}{ \textbf{Dec. 2000}}&
\tabularnewline
 B &
 25.86(0.03) &
 -0.18(0.04) &
 0.29(0.02) \tabularnewline
 V &
 25.94(0.04) &
 0.00(0.04) &
 0.19(0.03) \tabularnewline
 I &
 24.80(0.01) &
 -0.17(0.01) &
 -0.09(0.02) \tabularnewline
\multicolumn{3}{c}{ \textbf{Feb. 2002}}&
\tabularnewline
 V &
 25.69(0.01) &
 -0.01(0.01) &
 0.15(0.01) \tabularnewline
 I &
 24.59(0.02) &
 -0.03(0.01) &
 -0.06(0.01)  \tabularnewline
\hline
\end{tabular}
\end{table}

\begin{figure}[htbp]

\caption{Residual between standard magnitudes and calibrated magnitudes for
the Landolt stars in the Dec. 1999 run.}

\includegraphics[%
  width=8cm]{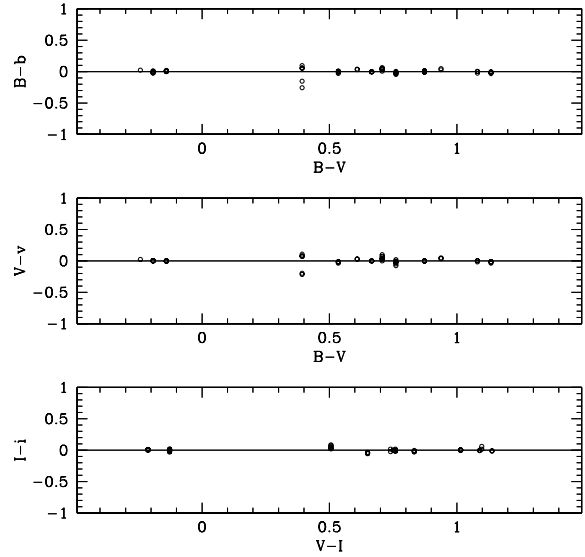} \label{cal99}
\end{figure}

\begin{figure}[htbp]

\caption{Residual between standard magnitudes and calibrated magnitudes for
the Landolt stars in the Dec. 2000 run.}

\includegraphics[%
  width=8cm]{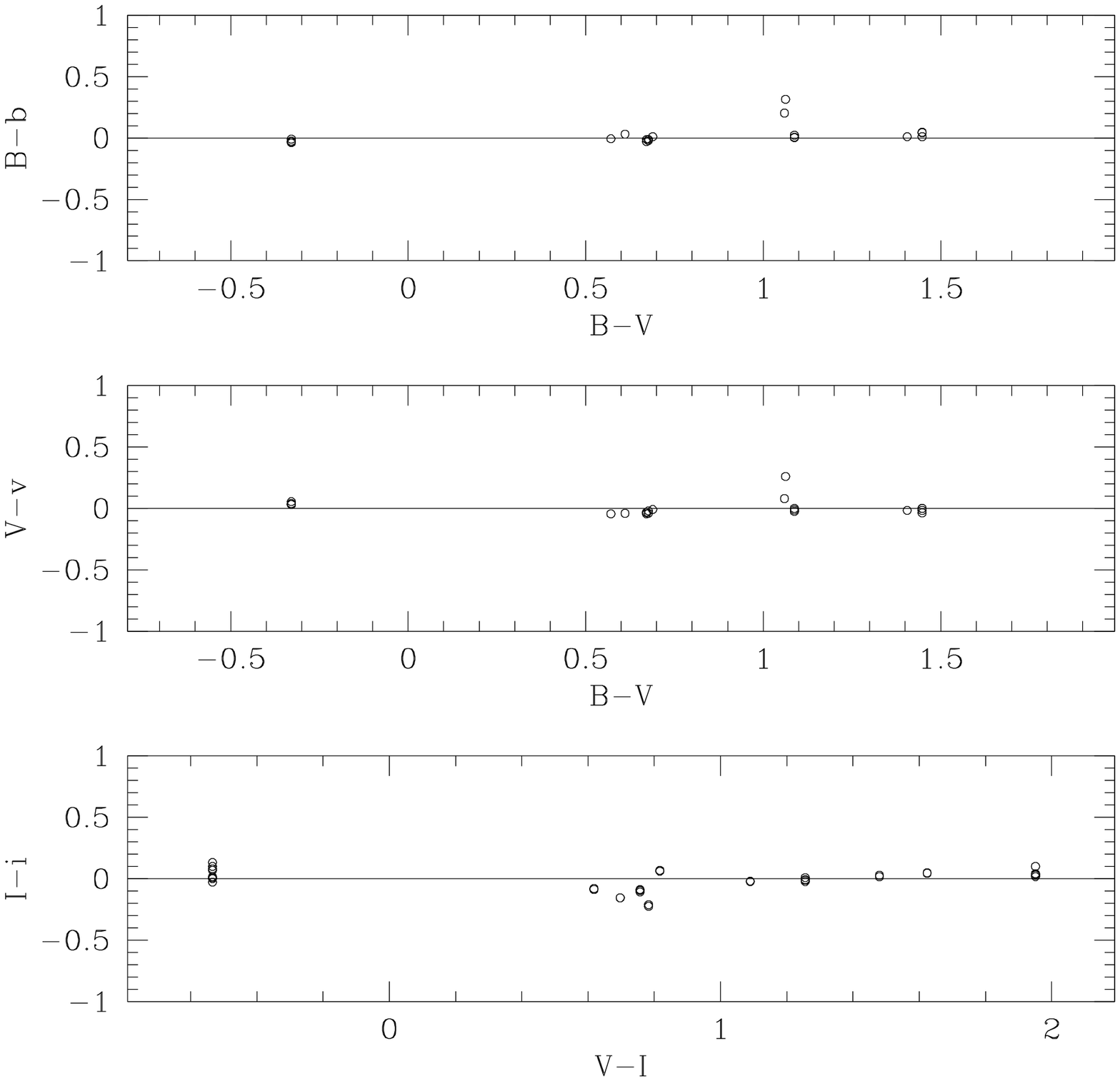} \label{cal00}
\end{figure}

\begin{figure}[htbp]

\caption{Residual between standard magnitudes and calibrated magnitudes for
the Landolt stars in the Feb. 2002 run.}

\includegraphics[%
  width=8cm]{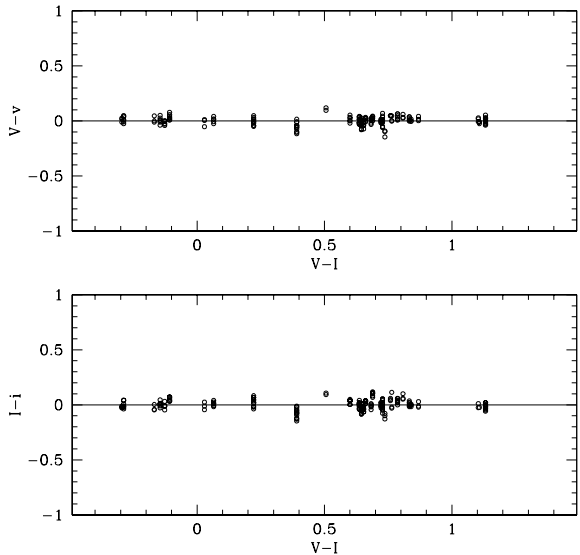} \label{cal02}
\end{figure}

Figs. \ref{cal99}, \ref{cal00} and \ref{cal02} show the difference
between standard mags. and calibrated mags. of the Landolt stars used
for calibrations in all the runs. As can be seen, the residuals
($\mathrm{Mag}_{std}-\mathrm{Mag}_{fit}$) are on the average very small 
and with no residual trends with the colors.

The HST-WFPC2 data-set has been calibrated following the procedure
describred in \citet{holtzmann95}. After applying the aperture corrections
and correcting for geometric distorsions and CTE effect, following
\cite{dolphin00}, we applied the published coefficients and calibrated
the data-set both in the {\footnotesize VEGAMAG} and {\footnotesize STMAG}
systems. While the V magnitudes are almost identical in the two systems,
the I magnitudes differ by a certain amount (around 1.2 mag, depending
on the WF chip).

The NTT data-set has then been used to report magnitudes in both system
to the ground NTT system, in order to perform also an internal check.

\section{The subdwarfs sample}

The sample of subdwarfs used for this study has been selected from
the HIPPARCOS list taken from the papers of \citet{gratton97} and
\citet{reid97}. The requirement was that the sample be as close as
possible in metallicity to the target cluster, in order to reduce
systematics introduced by any trasformation that would have been needed
to report the values to the metallicity of M~5, and be distributed
over a wide range of magnitudes and colors in order to overlap a sufficiently
large portion of MS and ensure, in this way, a more robust fit to
the cluster MS mean ridge line. The sample consists of 6 stars, 4
of which observed in the dec. 1999 run, the others in dec. 2000. Each
star has at least three measurements, in order to reduce the error
on the magnitudes and remove the effect of eventual bad pixels and/or
flat field dishomogeneities. 

Since these stars needed to be observed
with very short exposure times, the shutter delay of SUSI2 has been taken into
account. In particular, the typical exposure time was 0.5s, and the
shutter delay for SUSI2, taken from the SUSI2 web page at ESO
\footnote{http://www.ls.eso.org/Telescopes/NEWNTT/refdata/param\_ccd.html
shows that the exposure is uniform down to 0.3s. Since the stars were placed
close to the center of the chip, the resulting shutter effect turns out to be
negligible}.

Table \ref{subdwarfs} shows the calibrated
magnitudes and colors for the selected sample. The error indicated
is the internal photometric error, and is not representative of the
total error. To obtain this last, uncertainties on the reddening and
trigonometric parallax must be applied. As pointed out also by
\cite{gratton03}, these stars have virtually zero reddening. 
The error on the trigonometric parallax, given in the literature,
has been applied following the propagation of errors. The Lutz-Kelker
correction has been computed on the basis of the discussion by \cite{carretta00},
by using the formula adopted by the authors (Eq. 1 in the cited work), 
because the sample had the same selection effect as the one reported in that
paper. This value is of the order of 0.02 mag. The associated Malmquist
bias, for the same reason, is negligible.
By adding all the error sources, the average uncertainty increases to
$\sim0.03$mag but with a spread given by star-to-star differences. %

\begin{table*}[htbp]

\caption{The subdwarfs sample}

\label{subdwarfs} 

{\scriptsize }\begin{tabular}{lrrrrrrrlrc}
\hline 
Ident. &
 B &
 $\sigma$(B) &
 V &
 $\sigma$(V) &
 I &
 $\sigma$(I) &
 {[}Fe/H{]} &
 M$_{v}$&
 $\pi_{mas}$&
 Run \tabularnewline
\hline
HIP70681 &
 9.881 &
 0.001 &
 9.302 &
 0.001 &
 8.577 &
 0.001 &
 $-$1.09$\pm$0.07 &
 5.72$\pm$0.16 &
 19.16 &
Dec99 \tabularnewline
 HIP74234 &
 10.272 &
 0.001 &
 9.422 &
 0.001 &
 8.446 &
 0.001 &
 $-$1.28$\pm$0.07 &
 7.08$\pm$0.11 &
 33.68 &
Dec99 \tabularnewline
 HIP74235 &
 9.821 &
 0.001 &
 9.059 &
 0.001 &
 $-$&
 $-$&
 $-$1.30$\pm$0.07 &
 6.74$\pm$0.08 &
 34.14 &
Dec99 \tabularnewline
 HIP81170 &
 10.344 &
 0.001 &
 9.611 &
 0.001 &
 8.723 &
 0.001 &
 $-$1.14$\pm$0.07 &
 6.18$\pm$0.15 &
 20.71 &
Dec99 \tabularnewline
 HIP3026 &
 9.722 &
 0.001 &
 9.269 &
 0.001 &
 8.710 &
 0.001 &
 $-$1.50 &
 4.15 &
 9.57 &
 Dec00 \tabularnewline
 HIP24316 &
 9.924 &
 0.001 &
 9.448 &
 0.001 &
 8.838 &
 0.001 &
 $-$1.44$\pm$0.07 &
 5.28$\pm$0.15 &
 14.55 &
Dec00  \tabularnewline
\hline
\end{tabular}
\end{table*}

Since the selection has been done with the criterium on the metallicity
explained above and requiring that the objects be, of course, visible
from the site and in the period of observation, only six subdwarfs
have been selected, one of which saturated the I filter even with
the shortest possible exposure. The effect of metallicity on the subdwarf
fitting to the distance has been extensively discussed by several
authors in recent papers \citep[see,e.g.,][]{gratton97,carretta00,gratton03,reid97,pont98}
 and will not be readdressed here. However, it is worth recalling
that by selecting objects with a metallicity close to the cluster's
one, the uncertainities due to the metallicity spread are reduced,
and the correction to a common metallicity values is small. The
metallicity of the cluster has been recently recalibrated in the CG97
scale in $[\mathrm{Fe}/\mathrm{H}]=-1.10$, while ZW84, and subsequent
updates, give an estimate of $[\mathrm{Fe}/\mathrm{H}]=-1.40$. 
The metallicity scale of the subdwarfs, instead, has been recalibrated by
\cite{gratton03} and found to be, on average, 0.13$\pm$0.04 dex smaller than
the one given in \cite{carretta00}. This offset is to be taken into account 
in the DM determination, and will be discussed below.

\begin{table}[htbp]

\caption{Extra subdwarfs from literature.}

\label{added} 
{\scriptsize }\begin{tabular}{lrrrr}
\hline 
Ident. &
 M$_V$ &
 (V-I)$^a$&
 {[}Fe/H{]} &
 Source \tabularnewline
\hline
HIP57939 &
 6.61$\pm$0.02 &
 0.88 &
 $-$1.33$\pm$0.07 &
G03,H \tabularnewline
HIP57450 &
 5.59$\pm$0.25 &
 0.63 &
 $-$1.26$\pm$0.07 &
G97,H \tabularnewline
HIP74235 &
 6.74$\pm$0.08 &
 0.81 &
 $-$1.38$\pm$0.07 &
G03,H \tabularnewline
HIP79537 &
 6.84$\pm$0.02 &
 0.94 &
 $-$1.39$\pm$0.13 &
G97,H \tabularnewline
HIP7459 &
 5.43$\pm$0.23 &
 0.679 &
 $-$1.2 &
R01 \tabularnewline
HIP73798 &
 5.90$\pm$0.26 &
 0.727 &
 $-$1.2 &
R01 \tabularnewline
HIP89215 &
 6.50$\pm$0.26 &
 0.86 &
 $-$1.2 &
R01 \tabularnewline
HIP100568 &
 5.44$\pm$0.12 &
 0.678 &
 $-$1.00$\pm$0.07 &
R01,G03 \tabularnewline
\hline
\end{tabular}

~

\noindent$^a$ G97:Gratton et al. 97; H: HIPPARCOS catalog, for V-I color; \\
R01: Reid et al. 2001; G03: Gratton et al. 03.
\end{table}

\section{The Color-Magnitude Diagram}

\subsection{The NTT CMD}

\begin{figure}[htbp]

\caption{Ground CMD.}

\includegraphics[%
  width=8cm]{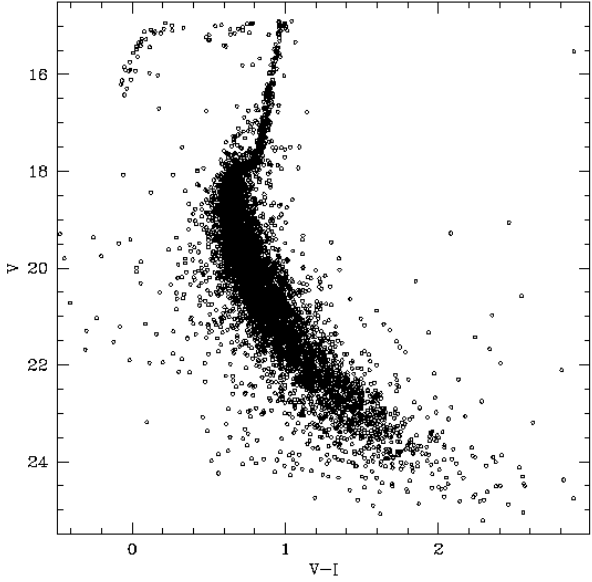} \label{groundcmd}
\end{figure}

\begin{figure}[htbp]

\caption{Run to run scatter in the ground measurements.}

\includegraphics[%
  width=8cm]{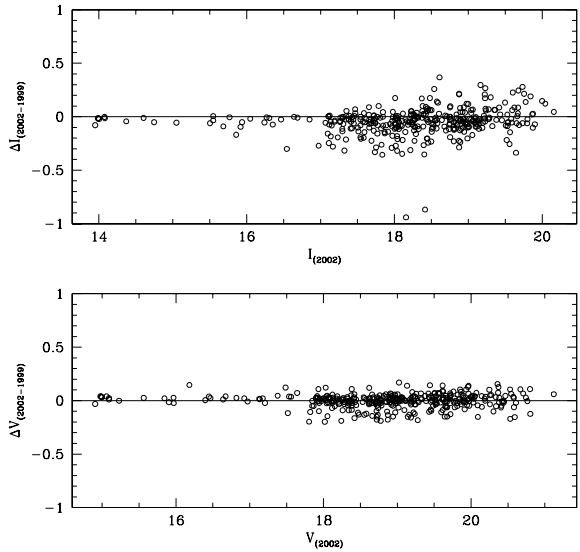} \label{figruntorun}
\end{figure}

The CMD derived from ground-based data (feb.~02) is shown in Fig.
\ref{groundcmd}. The diagram extends from the upper part of the RGB
down to about $V \sim 24$. A reasonably large magnitude range on the MS
is crucial in order to report the HST magnitudes (both {\footnotesize VEGAMAG}
and {\footnotesize STMAG}) to the ground NTT system. The CMD from
the dec.~99 run has been obtained under relatively poor seeing conditions
and hence came out to be not suited to perform an accurate calibration
of the HST data to the ground system. For this reason we adopted the
feb.~02 data-set after checking that the two sets of measurements
were self-consistent. Fig.\ref{figruntorun} shows the results of
the check.

\begin{figure}[htbp]

\caption{CMD from HST data.}

\includegraphics[%
  width=8cm]{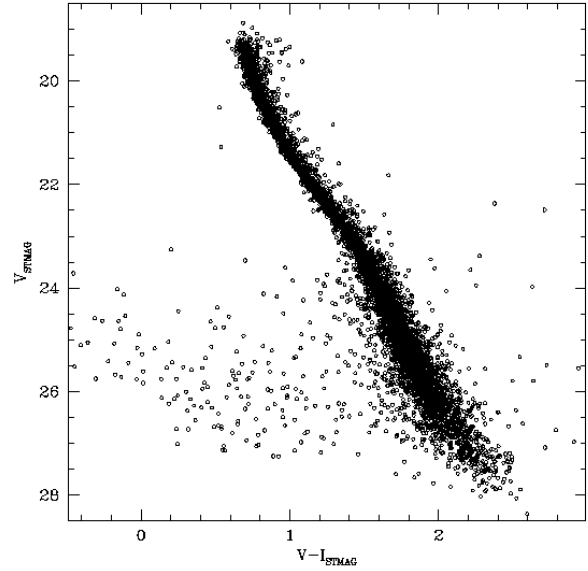} \label{hstcmd}
\end{figure}

\begin{figure}[htbp]

\caption{Final CMD.}

\includegraphics[%
  width=8cm]{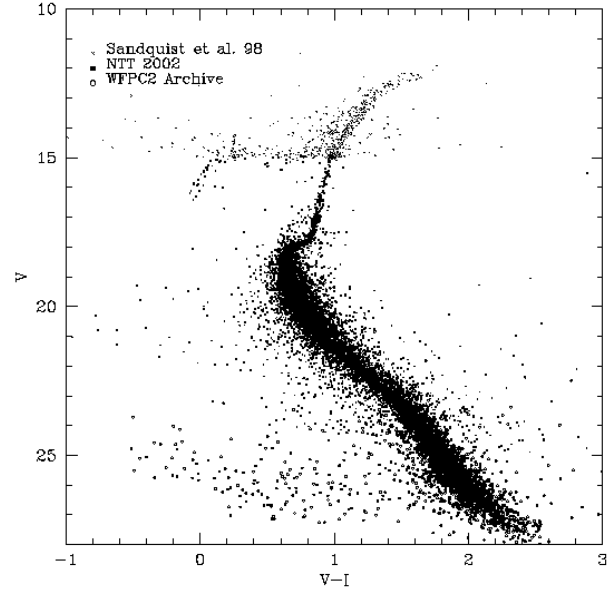} \label{finalcmd}
\end{figure}

\subsection{The HST CMD}

The HST measurements have been averaged, recording also information
on the number of frames in which each object was detected in each
filter. As final catalog we consider all the stars appearing in at
least 80\% of the frames (at least 5 V and 7 I). This choice is a
good compromise between taking all the measurements and taking only
the stars detected in all the frames. The first choice would contaminate
the CMD with all the spurious detection that appear by chance on only
one frame. The second leads to missing faint objects, like WDs. We remark 
the fact that the brighter part of the catalog (down to $\mathrm{V} \sim 26.5$)
is not altered by this selection criterion, as expected. The HST CMD
is reported in Fig. \ref{hstcmd}, where we show only the {\footnotesize VEGAMAG}
magnitudes. 

The whole M5 CMD is reported in Fig. \ref{finalcmd}. The HST data-set
extends approximately from $\mathrm{V}=19.5$ to $\mathrm{V}=28$, but the top
of the sequence is spread over because the chip is close to the saturation
and will not be considered in the following analysis. F555W and F814W
magnitudes have been reported to the ground based CMD via a relative
calibration equation of the form (for the V magnitude, but it applies
to the I magnitudes as well) 
$$V_{ground}=V_{HST}+zp+ct\times(V_{ground}-I_{ground})$$
The output for the V magnitude turned out to be identical for {\footnotesize STMAG}
as well as for {\footnotesize VEGAMAG}, while for the I filter we
derived, as expected, different transformations. The two HST magnitudes
systems have been checked for self-consistency after reporting them
to the ground NTT system. The two transformed magnitudes still show
some residuals that, over a range $20<\mathrm{V}<23$ is $\sim$0.01 mag, with
a shape that resembles that of a CMD. This is an indication that the
color effects were not completely eliminated with the transformations,
probably because of the larger scatter in the ground-based CMD and
the relatively small magnitude overlap of the ground- and space-based
data. However, even at the fainter magnitudes, the overall difference
between the two magnitudes is always below 0.05 mag 
(at $\mathrm{V} \sim 27.5$).
In the following, we will adopt as HST transformed magnitudes the
ones coming from {\footnotesize VEGAMAG}, because this system is closer
to the ground Vega system, so that the required transformation is
smaller and thus less prone to residual uncertainties%
\footnote{see also the {}``HST Data Handbook for WFPC2'' at \texttt{\footnotesize http://www.stsci.edu/hst/HST\_overview/documents/datahandbook}%
}.

\subsection{Mean Ridge Line}

\begin{figure}[htbp]
\caption{Comparison of ridge lines for M5.}
\includegraphics[%
  width=8cm]{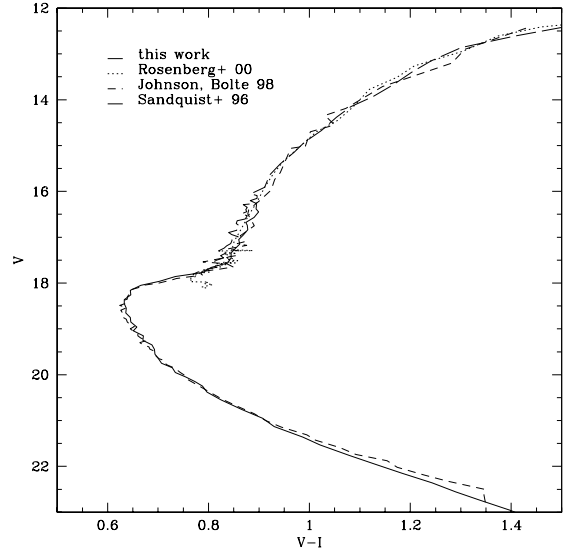} \label{compareridge}
\end{figure}

\begin{figure}[htbp]

\caption{The ridge line of the lower MS of M 5 with its uncertainty strip.}

\includegraphics[%
  width=8cm]{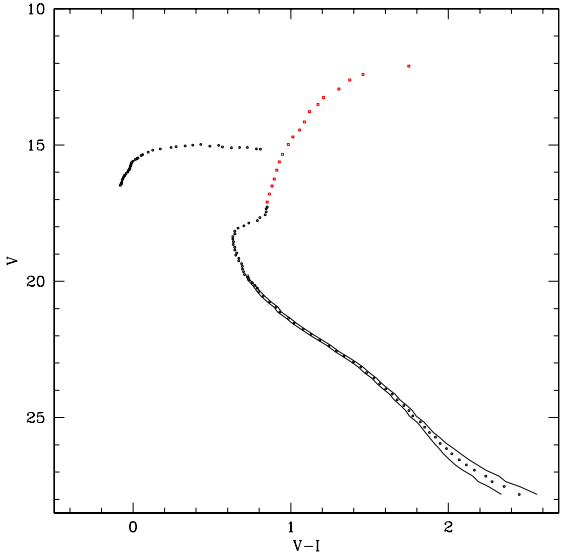} \label{ridgeline}
\end{figure}

The mean ridge line has been drawn on the faint part by using HST
data, on the upper MS until the sub-giant branch with the NTT plus
literature data, on the upper part of the RGB and the HB with the
\citep[R00]{rosenberg00} data, since their RGB is very narrow and well
populated. In general, in order to build the ridge line, the CMD has
been 'sliced' in magnitude bins and, for each bin, the median point
of the distribution has been determined, together with the associated
scatter. Only the objects with the lowest errors have been selected,
in order to ensure that the uncertainty on the ridge be small. The
zones with a strong curvature, i.e. at the bottom of the SGB, were
treated differently, by applying a polynomial fit to that range and
projecting the stars along the curved abscissa, then re-fitting iteratively
the polynom. In this way, the CMD is sliced along the curve and it
is possible to find the median point with high accuracy also in the
most bent magnitude bins. Fig. \ref{ridgeline} shows the ridge line
of the lower MS with its uncertainty strip. The narrowness of the
uncertainty strip will be used later to constraint the fit to the
subdwarfs sequence. As a comparison term, Fig. \ref{compareridge}
shows the ridge line determined as described above, the fiducial by
S96 and the ones derived with the same method by using
the data of \cite{bolte98} and R00, that are in close
agreement with ours. The irregularities in the

ridge lines obtained with literature data, that can be seen at the
base of the RGB, are due to the small size of the magnitude bins chosen
to draw the fiducial. Since we are not interested in a careful analysis
of those ridge lines, but only show them for sake of comparison, we
did not apply any smoothing or more refined technique in order to
improve it. It is, however, worth noting that our fiducial line extends
about 5 magnitudes below previous works. This fact is crucial because,
at least in principle, allows us to check model predictions down to
the bottom of the MS, as well as to use very faint subdwarfs to constrain
the distance fitting, if any (see below).

\begin{figure}[htbp]
\caption{Map of the fields covered by this work and recent literature.}\includegraphics[%
  width=8cm]{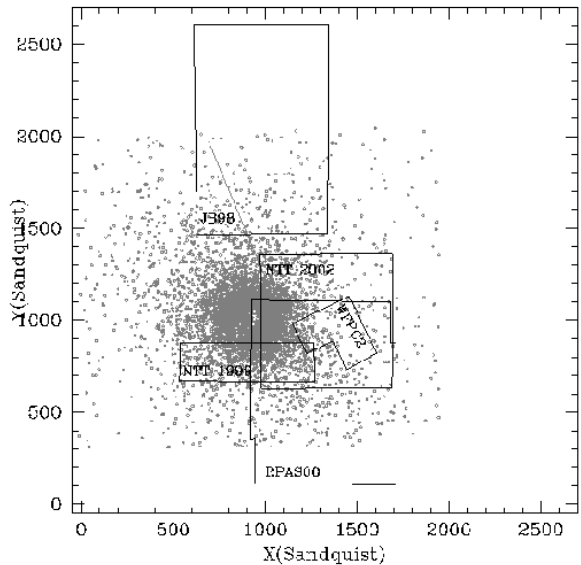} \label{maps}
\end{figure}

\begin{figure}[htbp]
\caption{Comparison with the data-set of S96}\includegraphics[%
  width=8cm]{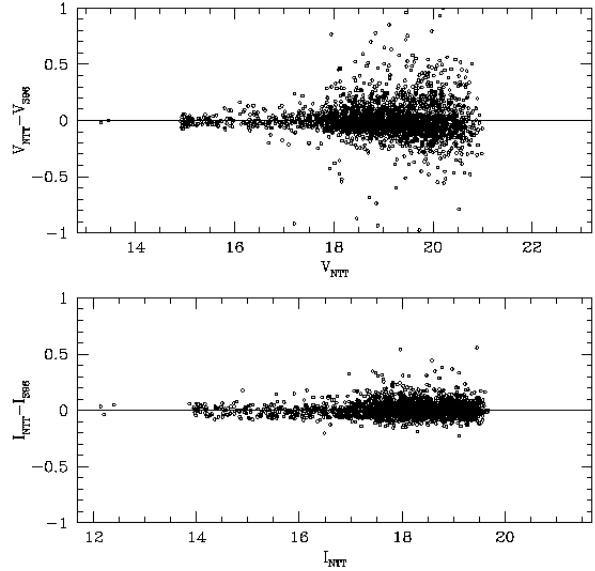}\label{sandquist}
\end{figure}

\begin{figure}[htbp]
\caption{Comparison with the data-set of R00}
\includegraphics[%
  width=8cm]{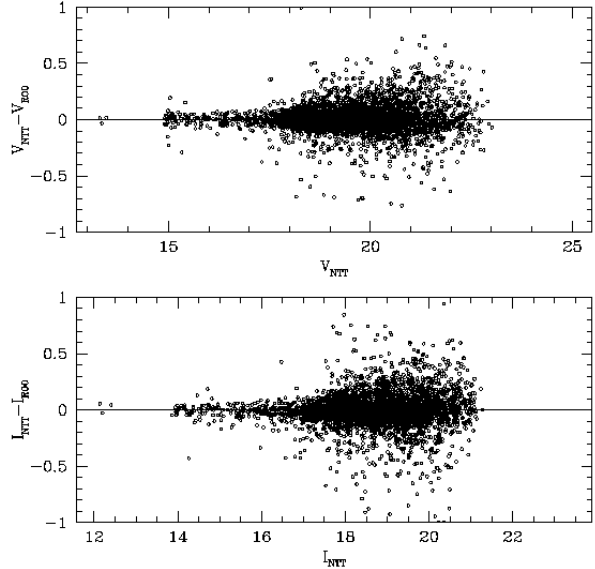}\label{rosenberg}
\end{figure}

\subsection{Comparison with previous photometry}

Beside the comparison of the ridge lines, a further check has been
done by comparing the NTT CMD with recent literature results, namely 
 R00, \cite{bolte98} and S96.
Fig. \ref{maps} shows the positioning of fields observed by us, the
HST pointing, and the three comparison works. The plots in
Figs. \ref{sandquist} and \ref{rosenberg}
show the comparison between our magnitudes and
the cited ones, where an overlapping of the fields permitted a direct
match of the objects. Our data show substantially a good agreement
with S96 and R00 but there are a few
issues that should be discussed: i) the average scatter is quite large,
meaning that a trend with the position could be present. Actually,
there is a small residual trend with position, with amplitude of the order 
of 0.02 mag. In the comparison with the two literature photometries
cited above the trend goes toward opposite directions, in the sense
that the trend along the X-coordinate, 
in the reference system of S96, resembles
the trend along the Y-coordinate in the R00 data-set, 
always in the reference system of S96 
(all the data have been reported to the system of S96), and
viceversa. The amplitude of the trend is larger with S96 than with
R00, and is absent if we compare the NTT 1999 data with both S96 and
R00. The only valid explanation for this is that a residual flat-field
is still present in some data-sets with respect to the others. In
fact, in the reduction of the Feb.~2002 data-set, we applied a more
refined technique for flat-fielding the data, described in \cite{hainaut98},
which takes into account also the lowest spatial frenquencies. The
1999 data-set has not been corrected for this effect. On the average,
this removed a residual trend in the fluxes. Since this procedure
has been applied with success in other works (Monelli et al. 2004,
in preparation), we find no reason to discard the procedure or re-calibrate
our data in the system of S96 or R00. However, the effect is visible
especially at fainter magnitudes ($\mathrm{V}>19$), while is almost absent
on the RGB. 

\begin{figure}[htbp]
\caption{Color-metallicity relation for the V-I color, taken from the models
 of \cite{scl97}.}\includegraphics[%
 clip,
  width=8cm]{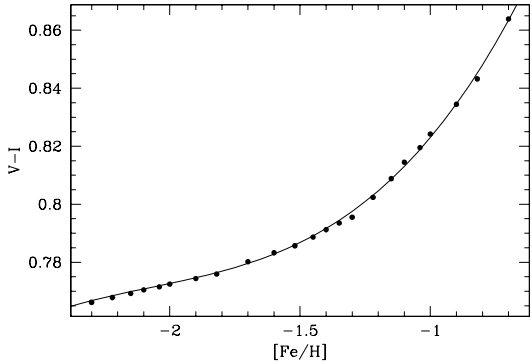}\label{colmet}
\end{figure}

\section{Distance via subdwarfs fitting}

\subsection{Error sources} 

  Before proceeding with the DM fitting, it is appropriate to
  discuss the contribution of the various error sources to the total
  uncertainty on the DM. \cite{gratton03} report in their Table
  1 a detailed list of all the error budget contributors before and after
  the re-analysis done in that paper. Taking that as a guideline, we can now
  see how the various uncertainties combine in our case. In Sect. 4 we have
  already mentioned some of the typical uncertainties in the subdwarfs
  analysis, and the fact that, since our sample have been drawn from
  the same lists published by \cite{gratton97,carretta00,reid97}
  and, partially, \cite{gratton03}, we expect that the uncertainty given by
  the Lutz-Kelker effect, the Malmquist bias and, clearly, the parallax, are
  essentially the same. Hence we adopt the values given in Table 1 of
  \cite{gratton03}.  
  
Photometric errors are very small both for the subdwarfs (see Table
\ref{subdwarfs}) and the cluster ridge line. For this last, in particular, the
very precise HST photometry coupled with a good ground-based calibration
allowed us to obtain an average error on the ridge line of the order of 0.005
mag at $(\mathrm{V - I}) = 0.7$. The error from the calibration relation
is also very small and of the order of 0.02 mag, since the photometric
conditions of the calibration night were close to excellent. In any case
subdwarfs and cluster have been observed in the
most homogeneous possible conditions and thus, 
the errors on the photometric calibration are expected to be very small. A
residual systematic trend could be present if the
calibrated magnitudes were not reported to the standard system, but the
obtained values are compliant with all the existing results, and thus we 
discard this possibility. 

Three other sources of error have to be now examined: the metallicity
scale, the cluster's reddening, the presence of
binaries. Their contribution affects both random and systematic error.

\subsubsection{Metallicity}

The metallicity scale is a major issue and has been widely
investigated in previous works \citep[see e.g.][]{gratton03,carretta00,reid97,pont98}. 
There are two main aspects to take into account: i) the spread in
metallicity given by the distribution in {[}Fe/H{]} of field stars;
ii) the relative metallicity scale used.
In order to reduce the first effect, the usual procedure is
to apply a relation that transforms tha values to a common metallicity 
(the cluster's).
In our case we decided to use only stars 
having metallicity values as close as possible to  M 5, so that the
transformation to a common metallicity is small. 
In order to estimate it,
  we used  the relation between $\mathrm{[Fe/H]}$ and
$\mathrm{(V-I)}$ at $\mathrm{M}_V = +6$, taken from the models of
  \cite{scl97}, that is shown in Fig. \ref{colmet} . 
A third order polynomial has been found to fit adequately the $(\mathrm{V-I})$
vs $\mathrm{[Fe/H]}$) relation in the range $-2.3 < \mathrm{[Fe/H]} < -0.5$.
The second effect stands on the systematic difference between
the two most commonly used scales, the ZW and the CG. 
\cite{gratton97} use an homogeneous metallicity
system, but our subdwarfs sample has been taken from that work and from
\cite{reid97}, who used the ZW scale. 
Since our cluster has metallicity estimates
$\mathrm{[Fe/H]}=-1.1$ in the CG scale, and $\mathrm{[Fe/H]}=-1.4$ in the 
ZW scale, we found it appropriate to report \cite{reid97} values to \cite{gratton97},
i.e. from ZW to CG, and then selecting stars from the \cite{reid97}
sample with the reported metallicity close to the cluster's one.

It should be noted that some of the stars in the present sample have not been
  recalibrated by \cite{gratton03}, hence a residual
  systematic spread might be present which affects the DM fitting as well, 
due to the co-existence of two slightly different recalibrations. 
If the value of
  0.13 dex is taken as average recalibration offset, we estimate a residual
  systematic error due to metallicity of $\pm0.008$ (at 1 $\sigma$) in the
  $(\mathrm{V-I})$. {\bf Given the slope of the MS between $\mathrm{M_V} = +5$
  and $\mathrm{M_V} = +6$, this propagates into an uncertainty of $0.04$ mag}.
The uncertainity on the metallicity (usually 0.1 dex) is, instead, added to
  the random error budget, and corresponds, following \cite{gratton03}, to
  $\pm 0.08$ mag.

\subsubsection{Reddening}

The cluster's reddening also plays a double r\^ole, since the associated error
affects the global random error and the uncertainty on the absolute amount of
reddening is a source of systematic error. We shall discuss quantitatively
this point below when we present the fitting procedure.

\subsubsection{Binaries}

The presence of binary stars affects both the
subdwarfs and the cluster sample. Although we selected bona fide single stars,
a residual quantity of binaries could still affect the subdwarf sample. 
We rely always
on the fact that the sample has been selected from the published lists of
\cite{gratton97} and \cite{reid97}, and thus assume their figure for the
contribution of subdwarf binaries to the total error, i.e. $\pm 0.02$
\citep[see][]{gratton03}. 
The effect of binaries on the cluster ridge line is slightly more
complicated. The effect of the presence of binaries in the cluster is to
spread the main sequence toward redder colors and, in extreme cases, to draw a
separate sequence, as has been detected in some open clusters 
\citep[see, e.g.,][]{marconi97}. 
In our case, the HST main sequence is very narrow, and a
well defined clear binary sequence would be visible. 
Since this is not the case, we tried
to estimate the effect of a MS spread due to an unresolved sequence of 
binaries. In order to do this, we
``rectified'' the main sequence and built a color histogram of the brightest
portion of the HST MS, that has a uniform width all over the considered range
$(20 < \mathrm{V} < 22)$. 
A gaussian profile has been fit to this profile, considering only
the blue tail of the distribution, that is in principle unaffected by the
presence of binaries. The fitted profile has then been subtracted 
from the original distribution. The
residual red tail is an indication of the binary contribution. This procedure
is quite robust and gives reliable figures for the uncertainties on the
position of the MS locus in the CMD. From this, we derive an estimate of a
possible systematic deviation of the MS ridge line from the ``true'' single
star MS given by binaries of {\bf $\sigma (\mathrm{V-I}) = 0.005$}, 
in addition to the random error due to the calculated spread of the MS. Again, 
this value in color propagates into a $\pm 0.025$ mag of systematic uncertainty. 
Table \ref{errors} below summarizes all the error sources involved in the 
DM fitting process.

\begin{table*}[htbp]

\caption{Contribution to the total uncertainty, in magnitudes, of the various
  error sources. Values are at 1$\sigma$ level.}

\label{errors} 

{\scriptsize }\begin{tabular}{lcc}
\hline 
 & Field Stars & Cluster \tabularnewline
\hline
Photometry & $\pm 0.001$ & $\pm 0.005 $ \tabularnewline
Calibration &  $\pm 0.02$ & $\pm 0.02 $ \tabularnewline
Lutz-Kelker &  $\pm 0.02$ & $-$ \tabularnewline
Malmquist   & Negligible & $-$ \tabularnewline
Parallax    &  $\pm 0.01$ & $-$ \tabularnewline
Metallicity (rand.) &  $\pm 0.08$ & $\pm 0.08$ \tabularnewline
Metallicity (syst.) &  $\pm 0.04$ & $\pm 0.04$ \tabularnewline
Reddening (rand.)  &  Negligible & $\pm 0.025 $ \tabularnewline
Reddening (syst.)  &  Negligible & $\pm 0.03 $ \tabularnewline
Binaries    &  $\pm 0.02$ & $\pm 0.025 $ \tabularnewline
\hline
\end{tabular}
\end{table*}

\subsection{Distance Modulus Fit}

\begin{figure}[htbp]
\caption{Fit of the cluster fiducial to the subdwarfs sequence. The stars with
 the label on the right belong to the observed sample. The others have been
 taken from published works.}\includegraphics[%
 clip,
  width=8cm]{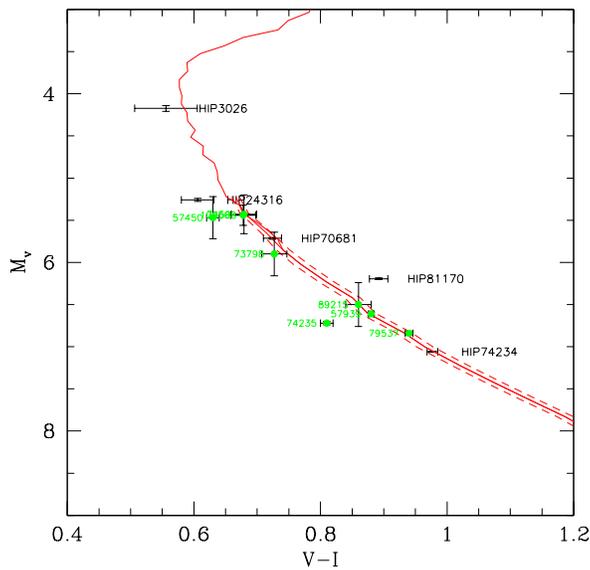}\label{fit_dm}
\end{figure}

By taking into account all the considerations and the caveats described
above, we used the subdwarfs sample to fit a DM to the
cluster data. In order to do this, we adopt a value of reddening of
$\mathrm{E(B-V)}=0.035 (\pm 0.005)$ \cite[and refs. therein]{carretta00}, corresponding
to $\mathrm{E(V-I)}=0.056$, {[}obtained from the relation in \cite{rieke85}:
$\mathrm{E(V-I)=1.6\times E(B-V)}${]}. {\bf The quoted uncertainty has been
  taken from the literature, and propagates into a $0.025$ uncertainty in 
  magnitude. The DIRBE maps \citep{schlegel98} report a reddening value in the 
  direction of M~5 of $\mathrm{E(B-V)} = 0.036$, consistent with the absolute 
  value given in \cite{carretta00}}. As shown in Fig. \ref{fit_dm}, the
brightest calibrating subdwarfs (HIP3026 and HIP24316) turn out to
be too blue to fit directly the cluster ridge line, because the cluster
MS has already evolved toward the turn-off point. However, they are useful
to draw the subdwarfs main sequence locus to be compared with our data.
Of the other three stars, HIP81170 is known to be a spectroscopic binary,
and this might justify its redder color. The
other two, HIP70681 and HIP74234, have the smallest error and have
then been used to fit the cluster ridge line.
We fit the cluster fiducial line
to the subdwarf sequence by shifting the fiducial according to the
reddening (in (V$-$I)!) and extiction -$3.1\times \mathrm{E(B-V)}$-, then with
a least square fitting of the difference between the used subdwarfs
and the cluster fiducial, with their errors.
Fig. \ref{fit_dm} shows the results of the best fit, that turns out to
be $\mu_0 = 14.44$. If a value of $\mathrm{E(B-V)} = 0.03$ is adopted, which
corresponds to $\mathrm{E(V-I)} = 0.048$, $\mu_0 = 14.41$ is obtained. The
error on the fit obtained from the combination of the random error sources
only is, in both cases, {\bf $\pm 0.09$, the uncertainty in the metallicity 
being the strongest contributor}. 
In order to increase the number of subdwarfs
suitable to form a fiducial reference sequence, we added to our sample
other subdwarfs listed in \cite{gratton97} and \cite{reid01}, having
$\mathrm{[Fe/H]}$ similar to that of M~5 and discarding known double stars. 
The added stars are listed in Table \ref{added} and shown with different
symbols (filled circles) in Fig. \ref{fit_dm}. It can be seen that most of the
added stars draw a good reference main sequence, with the exception of
HIP57450 and HIP74235, which are considerably bluer. On the
$(\mathrm{V},\mathrm{B-V})$ CMD, however, they lie on the MS, and thus we are
inclined to interpret their bluer colors as uncertainties in the I magnitudes,
that has been taken from the original HIPPARCOS catalog. For this reason, the
two outlying stars have been excluded from the fit. 

 It can be seen that the added stars define three well
separated sequences corresponding to the three metallicity ranges
selected, from the lowest (triangles, bluest), to the intermediate
(circles), to the highest (squares, reddest). 

With the extended sample of subdwarfs, our best estimate for the (dereddened) 
DM is $\mu_{0}=14.44$ if $\mathrm{E(B-V)}=0.035$
is adopted. If, instead, $\mathrm{E(B-V)}=0.03$ is adopted, as reported in
\cite{gratton97}, the DM turns out to be
$\mu_{0}=14.41$, thus confirming the value obtained with the
original, and homogeneous, sample.

The error associated to the fit above has been estimated from all
the error sources discussed above and reported in Table \ref{errors}, 
and considering random and systematic errors separately. 
From a statistical error propagation all over the involved
quantites, the uncertainty associated to the fit turns out to be $\pm0.09$ mag.
 A systematic uncertainty, combined from reddening and metallicity
scale, of $\pm0.07$ mag must then be added. {\bf The global uncertainty is
slightly larger than the one reported by \cite{gratton03}}, probably because 
the $\mathrm{V-I}$ color is more sensitive to metallicity variation and thus
the effect of variation in [Fe/H] and E(B$-$V) are amplified with respect to
the B-V color.

\section{Discussion and Conclusions}

This study was aimed at estimating the DM of M 5 through
the main-sequence fitting method, using as calibrating sequence a
set of local subdwarfs for which the trigonometric parallax from HIPPARCOS
was known. In order to achieve our goal we used the same setup telescope+instrument+filters
both for the cluster and subdwarfs data, namely the ESO-NTT equipped
with the camera SUSI2 and standards ESO V and I filters. 

Later, a set of subdwarfs from literature has been added to the original
sample in order to improve the sample size, and hence increase the
accuracy of the fitting. All the comparisons with recent data from
literature show that our data are fully consistent with previous 
determinations,
whose accuracy is known, both for the cluster and the local subdwarfs.
A final sample of $\sim$10 subdwarfs has been used to fit
a DM to the cluster. By taking into account the two
most commonly accepted values for the reddening (i.e. E(B$-$V)=0.03 and
0.035), we obtained two very close values for the (dereddened)
DM: $\mu_{0}=14.41\pm0.09\pm0.07$ and $\mu_{0}=14.44\pm0.09\pm0.07$,
respectively, where the first number comes from random error sources, the
second from systematic error sources. 
These values are in good agreement both with \cite{reid97}
($\mu_{0}=14.45$ obtained with the subdwarfs having the smallest error on the
parallax) and \cite{carretta00} ($\mu_{V}=14.59$ corresponding to
$\mu_{0}=14.48$), with a slight preference for the shorter value.
\cite{vdb02}, instead, found $\mu_{0}=14.38$ with the use of non-diffusive 
isochrones and $\mu_{0}=14.30$ with diffusive isochrones. 
In their work, they use a reddening value $\mathrm{E(B-V)} = 0.038$ and a
metallicity 
[Fe/H] $= -1.4$ for the cluster, 
discussing their preference for the ZW scale over the
CG  on the argument that the metallicity for M~5 is
inconsistent with the metal determination for the field stars, if the value is
as high as $-$1.1. Moreover, \cite{vdb02} choose to not report the field stars
to a common metallicity, but this effect acts on the spread of the relation
rather than on the absolute value.

In order to try to obtain a further constraint on the DM,
we tried to use a set of extreme subdwarfs, using an analogous approach
to \cite{reid98}, that has been found to be unapplicable because
both the main sequence locus at the faint magnitude level of the extreme
subdwarfs ($\mathrm{M}_{V} \sim 10$) is dispersed and the subdwarfs of intermediate
metallicity define a sequence dispersed as well. In order to reduce
the uncertainty, it would have been
desirable to have a set of subdwarfs of fainter magnitudes 
(i.e. below $\mathrm{M}_{V} \sim 7$) to take full advantage of the narrowness
of the HST main sequence, but they lack in the HIPPARCOS sample. Moreover, 
at such faint magnitude level, the lower main sequence stars are very cool and the
determination of their metallicity much more uncertain because
it is very difficult to take into account all the molecular species in the 
athmosphere at these low temperatures.

\begin{acknowledgements}
We wish to thanks Drs. Eric Sandquist and Jennifer A. Johnston for giving
us their published tables in electronic form and Dr. Olivier R. Hainaut
for useful and pleasant discussions. We would also like to thank the referee,
Dr. R. Gratton, for his useful comments that improved the presentation of this
paper. This work has been partially supported by the Italian Ministry of Education,
University and Research (MIUR) under grant ``COFIN 2000''.

\end{acknowledgements}

\end{document}